\documentclass[conference]{IEEEtran}
\IEEEoverridecommandlockouts
\usepackage{cite}
\usepackage{amsmath,amssymb,amsfonts}
\usepackage{algorithmic}
\usepackage{graphicx}
\usepackage{textcomp}
\usepackage{xcolor}
\def\BibTeX{{\rm B\kern-.05em{\sc i\kern-.025em b}\kern-.08em
    T\kern-.1667em\lower.7ex\hbox{E}\kern-.125emX}}

\usepackage{url}
\usepackage{multirow}
\usepackage{xspace}
\newcommand{\trace}{\textit{trace}\xspace}
\newcommand{\outcome}{\textit{outcome}\xspace}
\newcommand{\COMPS}{\textit{COMPS}\xspace}
\newcommand{\TESTS}{\textit{T}\xspace}

\newboolean{showcomments}
\setboolean{showcomments}{false}

\ifthenelse{\boolean{showcomments}}
  {\newcommand{\nb}[3]{
  {\color{#2}\small\fbox{\bfseries\sffamily\scriptsize#1}}
  {\color{#2}\sffamily\small$\triangleright~$\textit{\small #3}$~\triangleleft$}
  }
  }
  {\newcommand{\nb}[3]{}
  }

\newcommand\Roni[1]{\nb{\textbf{Roni:}}{orange}{#1}}

\usepackage{booktabs}
\usepackage{acronym}
\acrodef{ML}{Machine Learning}
\acrodef{LDP}{Learn, Diagnose, and Plan}
\acrodef{AI}{Artificial Intelligence}
\acrodef{DA}{diagnosis algorithm}
\acrodef{HP}{highest probability}
\acrodef{TP}{test planning}
\newcommand{\ML}{\ac{ML}\xspace}
\newcommand{\LDP}{\ac{LDP}\xspace}
\newcommand{\AI}{\ac{AI}\xspace}
\newcommand{\da}{\ac{DA}\xspace}

\newcommand{\tp}{\ac{TP}\xspace}
\newcommand{\conf}{\textit{conf}\xspace}

\begin{document}

\title{Learning Tests Traces
%\thanks{Identify applicable funding agency here. If none, delete this.} TODO-RONI
}
% \author{Eyal Hadad \and Roni Stern}

\author{\IEEEauthorblockN{Eyal Hadad} 
\IEEEauthorblockA{\textit{Software and Information Systems Engineering} \\
\textit{Ben Gurion University of the Negev}\\
Be'er Sheva, Israel \\
eyalhad@post.bgu.ac.il}
\and
\IEEEauthorblockN{Roni Stern} 
\IEEEauthorblockA{\textit{Software and Information Systems Engineering} \\
\textit{Ben Gurion University of the Negev}\\
Be'er Sheva, Israel \\
sternron@post.bgu.ac.il}
}

\maketitle

\begin{abstract}
Modern software projects include automated tests written to check the programs' functionality. The set of functions invoked by a test is called the \emph{trace} of the test, and the action of obtaining a trace is called \emph{tracing}. There are many tracing tools since traces are useful for a variety of software engineering tasks such as test generation, fault localization, and test execution planning. A major drawback in using  test traces is that obtaining them, i.e., tracing, can be costly in terms of computational resources and runtime. Prior work attempted to address this in various ways, e.g.,  by selectively tracing only some of the software components or compressing the trace on-the-fly. However, all these approaches still require building the project and executing the test in order to get its (partial, possibly compressed) trace. This is still very costly in many cases. 
% for  
% Many tools and research papers assume, either explicitly or implicitly, that the trace of every test and function is readily available. For example, popular test generation frameworks aim to maximize \emph{coverage} of the tests they generate, i.e., aim to maximize the set of functions in the traces of the generated tests. Test traces are also used by automated software troubleshooting techniques to localize observed bugs and to prioritize which tests to execute. 
% A main drawback in all of the above is that it requires knowing the trace of the generated or executed trace. To do so, one must execute the test, which can be costly in terms of computational resources. 
In this work, we propose a method to predict the trace of each test without executing it, based only on static properties of the test and the tested program, as well as past experience on different tests. This prediction is done by applying supervised learning to learn the relation between various static features of test and function and the likelihood that one will include the other in its trace. 
Then, we show how to use the predicted traces in a recent automated troubleshooting paradigm called \LDP, instead of the actual, costly-to-obtain, test traces. In a preliminary evaluation on real-world open-source projects, we observe that our prediction quality is reasonable. In addition, using our trace predictions in \LDP yields almost the same results comparing to when using real traces, while requiring less overhead. 
\end{abstract}


\section{Introduction}

% There are bugs 
The number of software projects and their size increase every day, while their time-to-market decreases. As a result, the number of bugs in software projects increase. Bugs damage the performance of software products and directly affect their customers. Therefore, 
software companies heavily invest in software quality and quality-related costs can consume as much as 60\% of the development budget~\cite{engel2007modeling}. 

% Tracing are useful for quality-related tasks
To maintain software quality, modern software projects include automated tests written to check the programs' functionality. The set of functions invoked by a test is called the \emph{trace} of the test. Many tools and research papers use test traces to perform a range of software engineering tasks, including test generation, bug isolation, and managing test execution. In test generation tools, traces are collected and used to compute \emph{coverage}, which is 
the union of the sets of functions in the traces of the generated tests. Maximizing coverage is a standard objective in popular test generation frameworks, such as EvoSuite~\cite{fraser2011evosuite}. 
In bug isolation, traces are used by software diagnosis algorithms such as Barinel~\cite{abreu2009spectrum} to localize the root cause of observed bugs. Traces are also used to prioritize which tests to execute~\cite{zamir2014using,elmishali2018artificial}. 

% Prior work 
A main drawback in using traces for all of the above tasks is that collecting traces is costly in terms of computational resources and runtime. This is because in order to obtain the trace of a test, one must build the project and execute the test while applying techniques such as byte-code manipulation to record its trace. All these activities can be very costly in real-sized projects, and the size of the resulting trace can be prohibitively large. Prior worked partially addressed this by compressing the trace while it is collected~\cite{reiss2001encoding,taheri2017parlot} and by choosing selectively which software components to include in a trace~\cite{tallam2007enabling,liblit2003bug,liblit2005scalable,zhao2017log20}. These approaches are very useful, but still require executing the test.

In this work, we propose to learn to \emph{predict} the trace of a test without executing it. This prediction relies only on static properties of the test and the tested program, as well as previously collected traces of other tests. \textbf{The first contribution of our work} is to define the trace prediction problem and model it as Binary classification problem. Then, we propose to use a \emph{supervised learning} algorithm over traces of previously executed tests to solve this classification problem, and suggest easy-to-extract features to do so. \textbf{This is the second contribution of this work}.

One of the benefits of having a trace prediction algorithm is that it can be used instead of real traces in software engineering tasks. We show this for the task of \emph{software troubleshooting}. In particular, we propose to integrate our test prediction algorithm in a recently proposed automated troubleshooting paradigm called \LDP~\cite{elmishali2018artificial}. \LDP aims to identify the root cause of an observed bug, and does so by using a combination of techniques from the Artificial Intelligence (AI) literature. It uses a software diagnosis algorithm to output candidate diagnoses. If this set is too large, \LDP uses a \tp algorithm to choose which tests to perform next in order to collect more information for the diagnosis algorithm. 
Prior work on the \tp component of \LDP assumed that the test planner knows the traces of the test it is planning to execute. We propose a simple \tp algorithm that can use the predicted traces instead of the actual, costly-to-obtain, test traces. \textbf{This is the third contribution of this work}. 

Finally, we perform a small-scale evaluation of our trace prediction algorithm and our \tp algorithm in \LDP on real-world open-source projects. Results show that while prediction quality can be improved, it is sufficiently accurate to be used by our \tp algorithm to guide \LDP to troubleshoot bugs almost well as  when using real traces.

\section{Background and Problem Definition}

% Definitions: an automated test, outcome, trace, software component
An \emph{automated test} is a method that executes a program in order to check if it is working properly. 
The outcome of running a test is either that the test has \emph{passed} or \emph{failed}, where a failed test indicates that the program is not behaving properly.\footnote{In general, there are automated tests that have other types of outcomes, e.g., tests that measure response time.} 
The \emph{trace} of a test is the set of software components, e.g., classes or methods, that were activated during the execution of a test. We denote by $\outcome(t)$ and $\trace(t)$ the outcome and trace, respectively, of a test $t$. 

% Tests and components of a given program. 
Modern software programs include a large set of automated tests. For a given program of interest, we denote its set of automated tests by $\TESTS$, and its set of software components by $\COMPS$. Note that for every test $t\in\TESTS$ it holds that $\trace(t)\subseteq \COMPS$.

% Current approaches to get the trace are based on executing it
There are several techniques for obtaining the trace of test after executing it. A common way to obtain the trace of a test $t$ is to modify the program's source code so that it records every function invocation, and then run $t$. 
For example, in Java programs such code modification can be done in runtime using byte-code manipulation frameworks such as ASM (\url{http://asm.objectweb.org}), 
BCEL (\url{http://jakarta.apache.org/bcel}), 
and SERP (\url{http://serp.sourceforge.net}).
Another way to obtain the trace of a test 
is to execute it with a debugging tool, and, again, record every function invocation. 
These tracing techniques have been used in practice in various tracing tools, such as iDNA~\cite{bhansali2006framework} and Clover (\url{https://www.atlassian.com/software/clover}). 
For a survey of tracing tools, see~\cite{alemerien2014examining}.

All these tools require running the test in order to obtain its trace. However, running a test can costly in terms of computational resources and time. Moreover, tracing techniques usually incur non-negligible overhead, e.g., for printing out the trace to a file. The goal of this research is to output the trace of an automated test $t$ without actually running it. We call this problem the \emph{trace prediction} problem, and apply supervised learning to solve it.
% Formally,
% \begin{definition}[The Trace Prediction Problem]
% For a given program and an automated test $t$,
% the trace prediction problem is to problem of 
% \end{definition}

\Roni{We should somehow refer somehow to symbolic execution, since it is also a technique to predict traces. It does so by analyzing the code itself}

\section{Learning to Predict a Trace}
\label{sec:learning}
% In this research, we intend to develop a tests trace prediction method and use it in order to improve automated software fault diagnosis. 

% Overview: ML
In this section, we provide relevant background in supervised learning and show how the trace prediction problem can be solved with supervised learning techniques.

\subsection{Supervised Learning}

% An instance, a label
Supervised learning is perhaps the most widely used branch of \ML. Broadly speaking, 
supervised learning aims to learn from example input–output pairs a function that maps from input to output~\cite{russell2016artificial}. 
Supervised learning is commonly used to solve \emph{classification} tasks. 
A classification task is the task of mapping a \emph{label} to a given \emph{instance}, where the set of possible labels is discrete and finite. 
A \emph{binary classification} task is a classification task in which there are only two possible labels.

% training set, features, how supervised ML works. 
To solve a classification task using supervised learning, one needs to accept as input a set of instance-label pairs, i.e., a set of instances and their label. This is called a \emph{training set}. Supervised learning algorithms  work, in general, as follows: they extract \emph{features} from every instance in the training set, and then run an optimization algorithm to search for parameters of a chosen mathematical model that maps values of these features to the correct label. 

% The output: a classifier
The output of a supervised learning algorithm is this mathematical model along with optimized values for its parameter. In the context of classification, this is called a \emph{classifier}. A classifier can be used to output a label for a previously unseen instance, by extracting the features of this instance, inserting their values to the learned mathematical model, and outputting the resulting label.

 \subsection{Trace Prediction as a Binary Classification Problem}

% We used binary classification in order to predict the test trace.
% Binary classification is the task of classifying the elements of a given set into two groups (predicting which group each one belongs to) based on a classification rule. In binary classification the object is to correctly divide the objects into 2 different groups. The data is divided into groups that does not always at the same size, this is specific case called imbalanced data. 

 % Overview
 %Next, we show how trace prediction is, in fact, a binary classification problem, and   then propose features that can be used to solve it with supervised learning algorithms. 

%To model trace prediction as a binary classification task, we define what is an instance, a label, and how to obtain a training set. 
Trace prediction can be viewed as a binary classification task.
An \emph{instance} in trace prediction is a pair $(t,c)$ where $t$ is a test and $c$ is a software component. The \emph{label} is true if and only if $c$ is in the trace of $t$, i.e., iff $c\in\trace(t)$. The corresponding binary classifier is a classifier that accepts a pair $(t,c)$ where $t\in \TESTS$ and $c\in \COMPS$, and outputs true if $c\in\trace(t)$ and false otherwise. We refer to such a classifier as a \emph{trace classifier}. A trace classifier can be used to solve the trace prediction problem: for a given test $t$, run over all software components $c\in \COMPS$, and return only the components labeled as true by the classifier.

To solve this binary classification problem with supervised learning, we need a training set of 
test-component pairs and their correct label. 
To generate such a training set, we propose to run a subset $T_{train}\subset \TESTS$ of all automated tests and record their trace. 
Generating this training set is costly. However, as we show in our experimental results only a small fraction of all tests is needed in order to obtain a sufficiently large training set. In addition, this training step can be done only once for every major software version, as opposed to every time one needs to obtain a trace of a test.

% Key factors: 
Given a training set, we can use an off-the-shelf supervised learning algorithm to learn a trace classifier. 
Key factors in the successful application of such algorithms are  
(1) which optimization algorithm to use,
 (2) which classifier model to choose, 
 and (3)  which features to extract from each instance.  
 There are many general-purpose optimization algorithms for supervised learning, such as stochastic gradient descent~\cite{bottou1991stochastic} and Adam~\cite{diedrik2015adam}. 
 Similarly, popular classifier models such as decision trees~\cite{quinlan1986induction} and forests~\cite{breiman2001random}, support vectors~\cite{chang2011libsvm}, and artificial neural networks~\cite{haykin2009neural}, are commonly used in supervised learning. Designing useful features, however, is often done manually, with the aid of a domain expert. 

\subsection{Feature for Trace Prediction}
\label{sec:features}
%Useful features are features that help the learning algorithm discriminate between instances of different labels. \
We extracted two types of features for trace prediction: features based on \emph{call graph analysis} and features based on \emph{syntactic similarity}. 

\subsubsection{Features based on Call Graph Analysis}

Running a test and extracting its trace is a form of \emph{dynamic code analysis}. %Since we want to predict the trace of a test without running it, we cannot use features obtained from dynamic code analysis. 
\emph{Static code analysis} is an alternative common form of computer program analysis in which the program's source code is analyzed without running the actual program. Thus, information extracted in this way is especially suitable for our purposes, since we aim to predict a test's trace without running it. 

% Call graph
Static code analysis tools output a variety of source code artifacts, such as code smells~\cite{van2002java} and code complexity metrics. \emph{Call graph} is a standard output of many static code analysis tools that is particularly useful for our purposes. It is a graph in which every node represents a function and there is a directed edge from $n$ to $n'$ 
if the function represented by $n$ contains a call the function represented by $n'$. Note that it does not mean that every invocation of $n$ will also invoke $n'$, e.g., when the call to $n'$ is in an ``\texttt{if}'' branch that was not reached.

Obviously, the call graph contains important information with respect to whether a function is in the trace of a test or not. For example, if there is a path in the call graph from a function $n$ to a different function $n'$, then there exists an invocation of $n$ in which $n'$ will be in its trace.\footnote{Strictly speaking, it may not be true, as there may be branches in a function that can never be used, e.g., ''\texttt{if (1!=1) do X}''. However, this is not common.} In practice, for a  given project, we generate its static call graph and extract the following features for a given test function $t\in\TESTS$ and $c\in\COMPS$:

\begin{itemize}
  \item \textbf{Path existence.} A Boolean feature that indicates whether there is a path between $t$ and $c$. 
  \item \textbf{Shortest path.} The length of the shortest path between $t$ and $c$. 
    \item \textbf{Target in-degree.} The in-degree of $c$. 
    \item \textbf{Source out-degree.} The out-degree of $t$. 
\end{itemize}
% Those features obtain the graph information in two different aspects:
% Information that explain the connection between two nodes in the graph, and information about the nodes them-self in the graph.
% The first two features, Path existence and Shortest path add information about the connection between the test and the method in the graph, where there is a complete correlation between the two in order to increase there influence in the learning phase.
% The last features, Target in-degree and Source out-degree add information about the test and method them-self without taking into account the connection between them in the call graph.
% Because test nodes don't have entry edges we choose to count there out edges, and count the entry edges for the method nodes.

%\Roni{TODO: Add some intuition behind each feature}
%Eyal - I have added.

%    \item \textbf{Number of paths.} A numeric feature that counts the number of paths How many path there is between the two components in length 1 to 10.

\subsubsection{Syntactic Similarity}

A major limitation of call graphs generated by static code analysis is that they cannot detect \emph{dynamic} function calls. These are function calls that occur e.g., due to function polymorphism or dynamically loaded libraries. Thus, a component $c$ may be in a trace of a test $t$ even if there is no path from $t$ to $c$ in the call graph. More generally, the call graph by itself does not provide sufficient information to determine 
the trace of a test. 

To partially fill this gap, we consider another type of features that is based on \emph{syntactic similarity} between 
the name of the test $t$ and the name of the component $c$. The underlying intuition is that if a test and a class have similar names, then it increases the chances that the $t$ aims to use $c$ and thus it will be in its trace. 

Based on this intuition, we extracted the following features: 
\begin{itemize}
    \item \textbf{Common words.} The number of words that appear in both the test's name and in the component's name. Names were split to words using Camel notation, i.e., a capital letter is assumed to mark the start of a new word. This notation is standard in Java and other languages. 
    
    \item \textbf{Name distance.} The edit distance between the test's name and the component's name, normalized to the [0,1] range by dividing the edit distance with the length of the longer name. %\Roni{Eyal, how did you normalize?}      Eyal - explained.
    %This features is intended compensate for the binary nature of the previous feature, where a word is either exactly the same and used in both names or not.
\end{itemize}

In our experiments, every test $t$ is a method in a JUnit test class $T$ and every component $c$ is a method in some class $C$. We computed the above syntactic similarity features for the names of $t$ and $c$, and for the names of $T$ and $C$. Thus, we extracted four syntactic similarity features: class common words, method common words, class name distance, and method name distance.

% \subsection{Learning Algorithms}

% We tried 3 different algorithms for this classification task: SVM, Random Forest, Neural Network.
% The neural network manages to deal with the imbalanced data in a better way so we used it in our experiment. 

\section{Troubleshooting with Predicted Traces}
\label{sec:troubleshooting}
In this section, we present how our trace prediction algorithm can be used for \emph{automated software troubleshooting}. In particular, we show how our trace prediction algorithm can be integrated in \LDP~\cite{elmishali2018artificial}, a newly-proposed \AI-based paradigm for \emph{software troubleshooting}. For completeness, we first provide relevant background on software troubleshooting and \LDP. 

\subsection{The Learn, Diagnose, and Plan (LDP) Paradigm}
% What is software troubleshooting 
Software troubleshooting is a process that starts with an undesirable behavior of a given software, and ends by isolating the software components that should be fixed to avoid this undesirable behavior. Concretely, we focus on the case where one or more automated tests fail, and aim to find the software components that should be fixed to make all tests pass.

\LDP is a paradigm for software troubleshooting that contains three \AI components: 
a \emph{\textbf{l}earning} algorithm, 
a \emph{\textbf{d}iagnosis} algorithm,
and a \emph{\textbf{p}lanning} algorithm. 

\subsubsection{Learning to predict bugs}
This \LDP component is fundamentally a \emph{software fault prediction} algorithm~\cite{catal2011software}. It automatically matches between previously reported bugs and the code revisions made to fix them to create a dataset of buggy software components. This dataset is used to train a classifier that accepts a software component and outputs whether that component is expected to have a bug or not. Such learning-based software fault prediction algorithms work surprisingly well, at least for mature programs~\cite{malhotra2015systematic,elmishali2019debguer}. 
In addition, most learning-based software fault prediction algorithms can output a real value that represents the \emph{confidence} that a component is faulty.

\subsubsection{Diagnosing software bugs}
% Diagnosis algorithms, diagnosis likelihood
This \LDP component is a \emph{software diagnosis} algorithm. 
A software diagnosis algorithm, also known as software fault localization, accepts a set of executed tests and their outcomes, and outputs one or more \emph{diagnoses}. Each diagnosis is a set of software components that may have caused the observed bug. 
There are many approaches for software diagnosis, for a survey see~\cite{wong2016survey}. 
The \LDP diagnosis component is based on Barinel~\cite{abreu2009spectrum}, a state-of-the-art software diagnosis algorithm. Barinel follows a \emph{spectrum-based fault localization} (SFL) approach. It accepts as input a set of tests that were executed, the outcomes of these tests, and their traces. 
Then, it runs a fast hitting-set algorithm to 
find minimal sets of components that ``hit'' the trace of every failed test. Every such set is returned as a diagnosis.

Barinel can scale to real-world software systems and has many extensions~\cite{elmishali2016data,perez2018leveraging}. However, it tends to output a large set of diagnoses, denoted $\Omega$. 
While each of these diagnoses is a set of components that \emph{may} have caused the bug, only one diagnosis is correct, i.e., contain the actual components that caused the bug. Thus, returning a large set of diagnoses is not useful. To mitigate this, Barinel outputs for every diagnosis $\omega\in\Omega$ also a score, denoted $p(\omega)$, that roughly correspond to the likelihood that the components in $\omega$ are indeed the root cause of the failed tests. In \LDP, this score function is modified so that it is affected by the fault likelihood estimates from \LDP's learning component~\cite{elmishali2018artificial}. Nevertheless, in many cases too many diagnoses are still returned that have similar, non-negligible, scores. 

% the a Bayesian combination 

% Prior work showed how the learning component in \LDP can improve the score $p(\omega)$ returned by Barinel to provide a more accurate estimate of the likelihood that $\omega$ is correct. 
% Even this \emph{data-augmented} variant of Barinel often outputs too many diagnosis that have similar, non-negligible, scores. 

\subsubsection{Planning additional tests}
This \LDP component is a \emph{planning algorithm}. 
The input to this planning algorithm is (1) the set of tests that were not executed to so far and $T_{left}\subset \TESTS$ (2) the set of diagnoses $\Omega$ that was returned by the \da, along with their scores. Its output is the next test to execute. 

In \LDP, the selected test is executed, obtaining its trace and outcome. This information is added to the set of tests given to \da, which outputs a potentially more refined set of diagnoses. If this set is still too large to be useful, the \tp algorithm chooses an additional test to be executed. This process continues until either the correct diagnoses has been found, or all tests in $\TESTS$ have been executed (i.e., $T_{left}=\emptyset$), or the time allowed for automated troubleshooting has expired. 

\begin{figure}
    \centering
    \includegraphics[width=\columnwidth]{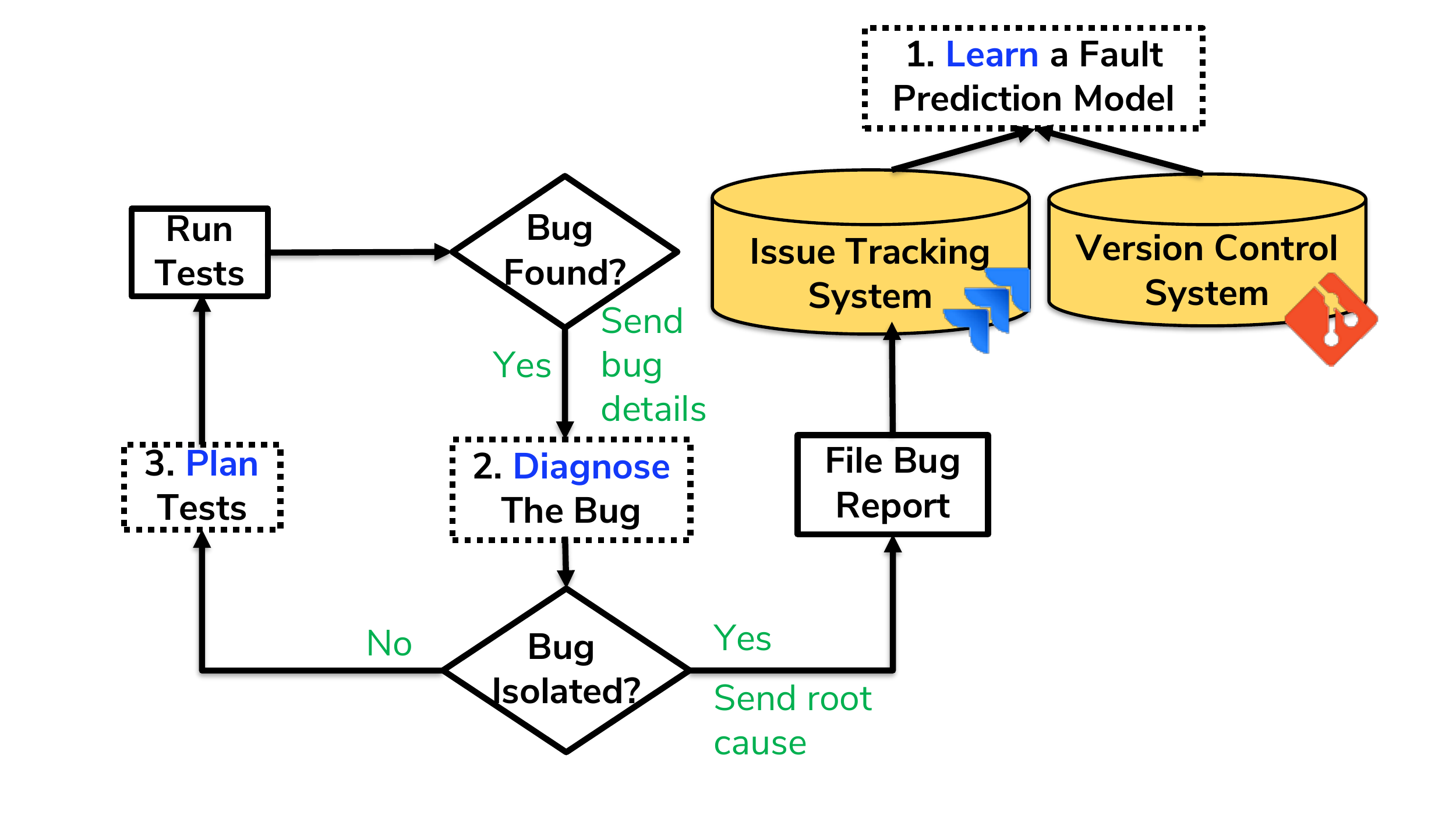}
    \caption{An illustration of the \LDP paradigm. Dashed rectangles are the AI components.}
    \label{fig:ldp}
\end{figure}

Figure~\ref{fig:ldp} illustrates the \LDP workflow. In the background, the learning component (dashed rectangle 1) learns a fault prediction model by extracting information about bugs that were previously reported in the issue tracker (JIRA) and their fix was committed in the version control system (Git). Then, when a test fails, the diagnose component (dashed rectangle 2) runs a \da and outputs a set of diagnoses. If needed, the planning component chooses which text to execute next (dashed rectangle 3).

\subsection{Test Planning with Predicted Traces}

% Other test planning algorithms
Several \tp algorithms have been proposed for troubleshooting~\cite{zamir2014using,elmishali2018artificial}, including sophisticated algorithms that compute information gain of each tests, or aim to solve an exponentially large Markov Decision Problem (MDP). 
These algorithms are computationally expensive, and, importantly, require knowing the trace of every test in $T_{left}$ although it has not been executed.

This is, of course, not realistic in practice. One approach to address this is to store the traces of every executed test. This has several limitations. First, it requires storing large amounts of data. Second, it requires running all tests at least once while recording their trace. This is very time consuming. Third, the trace of a test may change after modifying some component, and thus stored traces may be outdated and incorrect.

% Our approach
We propose a novel \tp algorithm that uses a trace prediction algorithm instead of knowing the actual trace of every test. This \tp algorithm requires a trace classifier, which we assume is generated using the trace prediction algorithm described in Section~\ref{sec:learning}. 
In addition, we require that the trace classifier output a \emph{confidence} score that roughly approximates the likelihood that the classifier is correct. As mentioned earlier, most learning-based classifier can output such a value. In our case, an instance is a component-test pair $(c,t)$ and the label is true iff $c\in\trace(t)$. Thus, the confidence of a trace classifier  approximates the likelihood that $c$ really is is in $\trace(t)$. We denote this confidence score by $\conf(c,t)$.

Our \tp algorithm also relies on computing the \emph{health state}~\cite{stern2015many} of every component $c\in\COMPS$ for the set of diagnoses and their scores returned by the \da.  
The health state of $c$, denoted $H(c)$, is defined as 
the sum of the scores of every diagnosis $\omega\in \Omega$  in which $c$ is assumed faulty. Formally, 
\begin{equation}
    H(c)=\sum_{\omega\in\Omega|c\in\omega}p(\omega)
\end{equation}
If the score of a diagnosis is indeed its probability to be correct and $\Omega$ is the set of all diagnoses, then $H(c)$ is the probability that $c$ is faulty~\cite{stern2015many}.

Finally, we are ready to define our \tp algorithm. It computes for every test $t$ the following utility function $U(t)$:
\begin{equation}
    U(t)=\sum_{c\in \COMPS} \conf(c,t)\cdot H(c)
\end{equation}
This utility function aims to approximate the expected number of faulty components in $\trace(t)$. Our \tp algorithm returns the test with the highest utility. %This is intended to aim for more failed tests, as in general, most test passsguide the e motivation for behind this is 
For practical reasons, in our implementation we computed the utility of a trace $t$ by only summing only the term $\conf(c,t)\cdot H(c)$ for the 40 components that are most likely in $\trace(t)$ according to our trace classifier. 

%For practical reasons, we computed the utility of every test $t$. 

\begin{figure}
    \centering
    \includegraphics[width=\columnwidth]{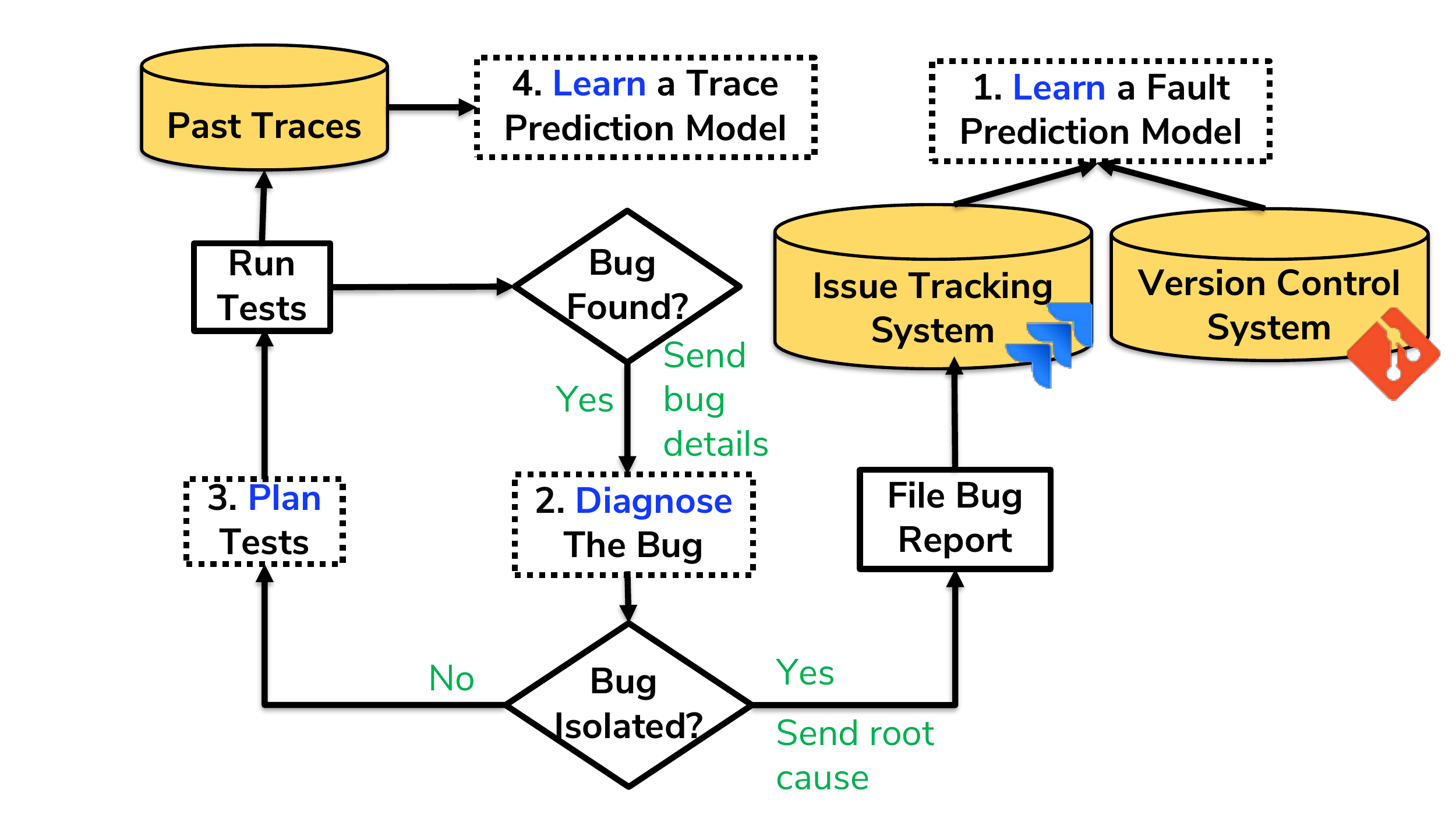}
    \caption{An illustration of the \LDP paradigm with our trace prediction added to the learning component. Dashed rectangles are the \AI components.}
    \label{fig:ldp-predict}
\end{figure}

Figure~\ref{fig:ldp-predict} shows how our trace prediction integrates well in \LDP. Whenever a test is executed and its trace is collected, this is added to the training set used to train our trace classifier (dashed rectangle 4). The \tp algorithm (dashed rectangle 3) can then use it as described above.

\section{Experimental Results}
In the first part of this section, we evaluate experimentally the performance of our trace prediction algorithm (Section~\ref{sec:learning}). In the second part of this section, we evaluate experimentally the performance of our \tp algorithm, which uses trace prediction, in the context of software troubleshooting (Section~\ref{sec:troubleshooting}). % and the performance of \LDP when using our \tp algorithm, which uses the trace prediction (Section~\ref{sec:troubleshooting}). 
% This evaluation 

% First, we provide general s
% In particular, we report on two sets of experiments we performed:
% \begin{itemize}
%     \item \textbf{Prediction experiments.} In this set of experiments, we run and evaluated our trace prediction algorithm, described in Section~\ref{sec:learning}. 
%     \item \textbf{Automated troubleshooting experiments.} In this set of experiments, we evaluated the performance of \LDP when using our \tp algorithm, which relies on the trace prediction algorithm. See ~\ref{sec:troubleshooting} for details. 
% \end{itemize}

%\subsection{Benchmark Projects}

\begin{table}
\label{tab:benchmark}
\caption{General statistics for the projects used in our experiments.} %Roni{TODO: Eyal, please verify this data - all the table}} Eyal - it's good now
\centering
\begin{tabular}{lrr}
\toprule
Project                     & Lang     & Math      \\ \midrule
First release                     & 2002 & 2004 \\
Last release                & 2018 & 2018 \\
Average number of functions & 4,000     & 7,400      \\ 
Average number of tests     & 1,660     & 3,500      \\ 
Test average length        & 11       & 39        \\ 
%Number of bugs              & 112       & 48        \\ 
% Imbalanced ratio            & 0.28\%  & 0.53\%   \\ this can be computed by dividing test length and number functio
%bugs source                 & BugMiner & Defects4J \\ \hline
\bottomrule
\end{tabular}
\end{table}

For our experiments, we used two popular open source projects: Apache Commons Math and Apache Commons Lang. Both projects are Java projects and use the Git version control system. Table~\ref{tab:benchmark} lists general details about these projects, including the date of the first and last release, the average number of functions and tests in a version, and the average length of a trace of a test. %, and the number of reported bugs we collected (see details below). 
%,  and the average percentage of all functions in the project that are used in given test (the ``Imbalanced ratio'' row).
%and the project was created, when it was last updated, number of contributors to the project, versions, classes, and methods. 

\subsection{Trace Prediction Experiments}
\Roni{Eyal - how many versions were there for every project?} 
% What did we do with each project
For each project, we chose a set of versions. 
For each version, we created a dataset of labeled component-test pairs by executing all available automated tests and collecting their traces. To speedup this process, we limited our tracing to 10\% of the methods. 
Thus, for a version with 100 tests and 2,000 methods, we obtained a dataset with 20,000 instances. This dataset was split equally (50/50) to train and test sets.

% Prediction algorithms we used
\subsubsection{Learning algorithms} For learning, we used a feed-forward artificial neural network 
and trained it using back-propagation. There are many possible neural network architectures, solvers, and activation functions. The configuration we used is a single hidden layer with 30 neurons, the Adam optimization algorithm~\cite{diedrik2015adam}, the ReLu activation function~\cite{nair2010rectified}, and a maximum of 3,000 iterations for training. We also tried other configurations, including different optimization algorithms (Newton's method and stochastic gradient descent) and activation function (Sigmoid), but the above configuration worked best. In addition, we tried a deep neural network architecture that includes 5 hidden layers, each with 30 neurons. We refer to the first architecture as ``NN'' and to the deep 5-layers architecture as ``DNN''.

% Using Adam as gradient descent optimization algorithm instead of sgd or quasi-Newton. made the biggest difference at both running time and performance. The AUC increased at 10 percentage in avg, and the training time reduced significantly.
%[[TODO: Details on the networks: how many layers, how many neurons, .... source code]]

\subsubsection{Evaluation metrics} 
We used standard supervised learning metrics to evaluate our trace prediction algorithm. Specifically, we report the following metrics. 
\begin{itemize}
 \item \textbf{True negatives (TN).} Percentage of instances where the target function is not in the trace and it is classified correct as such.
 \item \textbf{False positives (FP).} Percentage of instances where the target function is not in the trace but it is incorrectly classified as in the trace. 
 \item \textbf{False negatives (FN).} Percentage of instances where the target function is in the trace but it is incorrectly classified wrong as not in the trace. 
 \item \textbf{True Positives (TP).} Percentage of instances where the target function is in the trace and it is classified correctly as such.
 \item \textbf{Area under the receiver operating characteristic curve (AUC).} The receiver operating characteristic (ROC) curve plots the TP rate against the FP rate for different threshold values. A perfect classifier has an AUC value of 1 and a random classifier has an AUC of 0.5. %this metric is used in classification analysis in order to determine which of the used models predicts the classes best.\Roni{TODO-EYAL Explain. You can copy from your thesis proposal} Eyal - done.
  \item \textbf{Acc.} Accuracy is the ratio of instances that are classified correctly. This ratio is given by $\frac{TP+TN}{TP+TN+FP+NF}$. 
%tp + tn / tp + tn + fp + fn.
%  In other words it describes how many times the classifier classify right. \Roni{TODO-EYAL Explain. You can copy from your thesis proposal}
%  Eyal - done.
 \end{itemize}
All of these are standard machine learning metrics, commonly used to evaluate binary classifiers. For a more comprehensive discussion on these metrics, see~\cite{mitchell1997machineLearning}.

\begin{table}
\caption{Results of the prediction experiments.}
\label{tab:prediction}
\resizebox{\columnwidth}{!}{
\begin{tabular}{@{}lccccccc@{}}
\toprule
Algorithm & Project & AUC     & Acc     & TN      & FP     & FN     & TP  \\ \midrule
NN        & Lang    & 0.795 & 0.998 & 99.68\% & 0.07\% & 0.10\% & 0.15\%  \\
DNN       & Lang    & 0.779 & 0.998 & 99.69\% & 0.04\% & 0.12\% & 0.15\%  \\
NN        & Math    & 0.602 & 0.993 & 99.28\% & 0.27\% & 0.35\% & 0.10\%  \\
DNN       & Math    & 0.596 & 0.996 & 99.52\% & 0.03\% & 0.36\% & 0.09\%  \\ \bottomrule
\end{tabular}
}

\end{table}

\subsubsection{Results} Table~\ref{tab:prediction} presents the results of our prediction experiments. 
Note that the results in every data cell is the average obtained by our algorithm over all datasets.

% Acc. good
The accuracy obtained for both architectures (NN and DNN) is very high. For example, in the Math project the average accuracy is 0.993 and 0.996 for NN and DNN, respectively. However, these high accuracy results are misleading, because our dataset is highly \emph{imbalanced}. That is, most methods are not part of a given trace, and thus the true label of most instances (i.e., most method-test pairs) is negative. 
For example, according to Table~\ref{tab:benchmark} there are 4000 functions in Lang and the average number of functions called in a test is 11. Thus, less than 0.3\% of all instances are expected to be positives. This means that a naive trace classifier that says for every pair ($c,t$) that $c$ is not in the trace of $t$, which be quite accurate.

% AUC bad 
 AUC is a common metric to evaluate binary classifiers over imbalanced datasets. Indeed, we observe that the AUC for NN and DNN is far from perfect. 
For example, in the Lang project, the AUC of NN and DNN is 0.795 and 0.779, respectively, while the AUC of NN and DNN is only 0.602 and 0.596, respectively. %For reference, note that perfect classifier has an AUC of 1 and a random classifier has an AUC of 0.5. 
We hypothesize that it is possible to increase the AUC by incorporating known techniques for handling imbalanced datasets, such as under-sampling and over-sampling~\cite{chawla2002smote}, as well as devising more sophisticated features. However, as we show in the next set of experiments, our trace prediction was accurate enough to guide software troubleshooting effectively.

\begin{table}
\centering
\caption{Feature importance in the Lang project.}
\label{tab:feature-importance}
\begin{tabular}{@{}lcc@{}}
\toprule
Feature name          & NN   & DNN  \\ \midrule
Shortest path            & 0.10 & 0.16 \\
Target in-degree      & 0.02 & 0.00 \\
Path existence        & 0.12 & 0.00 \\
Method common words & 0.00 & 0.02 \\
Method names distance      & 0.04 & 0.00 \\ \bottomrule
\end{tabular}
\end{table}

% Features
To gain a better understanding of the impact of the different features (detailed in Section~\ref{sec:features}), we performed an \emph{all but one} analysis. 
An all but one analysis works as follows. We choose a feature $f$ and train two models: 
one with all features and one with all features except $f$. Then, we evaluate the performance of both models on the test set. The importance of $f$ is computed by the difference between the AUC of the model that considers all features and
the AUC of the model that considers all features except $f$. 
Table~\ref{tab:feature-importance} shows the importance of each feature using this analysis, for the NN and DNN models. As can be seen, the most important features are those that are based on static code analysis, namely, the length of the path in the call graph from the test to the method (``Path length'') and whether such a path exists (``Path existence''). 
The features based on syntactic similarity -- common words and name distance -- are also useful, but to a lesser degree.

% DNN not helpful
Interestingly, we do not observe a significant difference between the shallow neural network (NN) and the  deep one (DNN). For example, the AUC of NN and DNN in the Lang is 0.795 and 0.779, respectively. Therefore, for the rest of our experiments, we used the NN. Note that the design space of constructing a DNN is very large and one may come up with a DNN architecture that would be more efficient. 
%\Roni{Maybe talk about the difference between NN and DNN}\Roni{Maybe summarize this part}

\subsection{Troubleshooting Experiment}

In the next set of experiments, we evaluated how our trace prediction can be used in \LDP, as described in Section~\ref{sec:troubleshooting}. 

% Evaluated algorithms
\subsubsection{\tp algorithms}
We compared our \tp algorithm, referred to as \emph{Predicted}, to two baselines. The first uses the real traces of the available tests. That is, it assumes that $\conf(c,t)$ is one iff $c\in\trace(t)$ and zero otherwise. 
We refer to this is \emph{Oracle}. 
The second baseline \tp algorithm we used, referred to as \emph{Random},  chooses a test randomly. 
Note that Oracle cannot be used in practice, since we cannot check if $c$ is in $\trace(t)$ without executing it. Thus, Oracle serves as an ``upper bound'' to the performance of our \tp algorithm, while Random  is expected to be worse than our \tp algorithm, serving as a ``lower bound''. 

\subsubsection{Evaluation metrics}
% Where did the bugs came from
To compare between these test planning methods, we run \LDP as described in Section~\ref{sec:troubleshooting} on a set of previously reported bugs from the Lang and Math projects. 
These bugs were collected from the Defects4j repository~\cite{defect4j} and by mining the issue tracking and version control systems of these projects. 
In total, we used 112 bugs for Lang and 48 bugs for Math. % (see Table~\ref{tab:general-statistics}). %\Roni{TODO Eyal: replace X and Y}. Eyal - done

% We run all tests and recorded their traces, and marked the traces that contain the buggy method as a failed test. \Roni{Not ideal, since tests may pass even if they contain a faulty method}
% Then, we chose a set of initial tests $T_{init}$ and verified that at least one of them is marked as a failed test. 

% Every bug was in a different code version
For each of these reported bugs we conducted the following experiment. First, we chose a set of initial tests and verified that at least one of them is marked as a failed test. A test is marked as a failed test if its trace contains the buggy function. Then, we run \LDP using one of the evaluated \tp algorithms  until one of the following conditions occur:
\begin{itemize}
    \item A diagnosis was found whose likelihood is above $S$, where $S$ is a parameter referred to as the \emph{score threshold}. We set $S=0.7$
     in our experiments.  
    \item More than $B$ tests have been run, where $B$ is a parameter, referred to as the \emph{test budget}. We set $B=50, 75, 100, 125,$ and $150$ in our experiments. 
\end{itemize}
If the troubleshooting process has stopped because of the first condition, we say it has \emph{converged}. 
Otherwise, we say that it has \emph{timed out}. 
Converging is desirable because it means the troubleshooting process believes it has successfully isolated the root cause of the bug. 

For each bug, we performed three experiments, one for each \tp algorithm. In each experiment, we recorded the number of tests performed until the troubleshooting process either converged or timed out. We refer to this as the number of \emph{steps} performed. Fewer steps is better, as it means less testing effort for troubleshooting. %the troubleshooting process believes it has successfully isolated the root cause of the bug while executing fewer tests. 

For out \tp algorithm, we used a trace classifier that was trained on an earlier version of the code. That is, when running an experiment on a bug reported for version $X$, we trained the trace prediction model on the dataset extracted for version $X-1$ or earlier. Also, recall that only 10\% of all methods were used to train the trace classifier.

% \begin{figure*}
%   \includegraphics[width=0.4\linewidth]{Images/Lang_bars.png}
%   \includegraphics[width=0.38\linewidth]{Images/Math_bars.png}
%  \caption{Number of steps until reached one of the stopping conditions for Lang (left) and Math (right) projects.} 
%  \label{fig:bars}
% \end{figure*}

\begin{figure}
\centering
  \includegraphics[width=0.75\columnwidth]{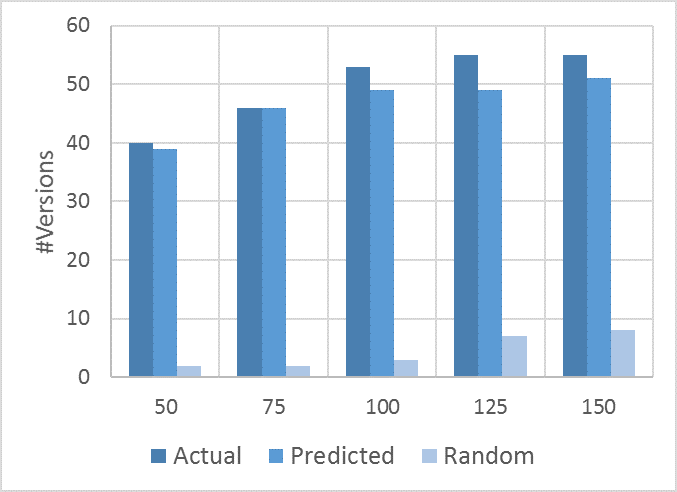}
 \caption{Lang project. \# of versions that converged for different test budgets.}
 \label{fig:bars-lang}
\end{figure}

\begin{figure}
\centering
  \includegraphics[width=0.75\columnwidth]{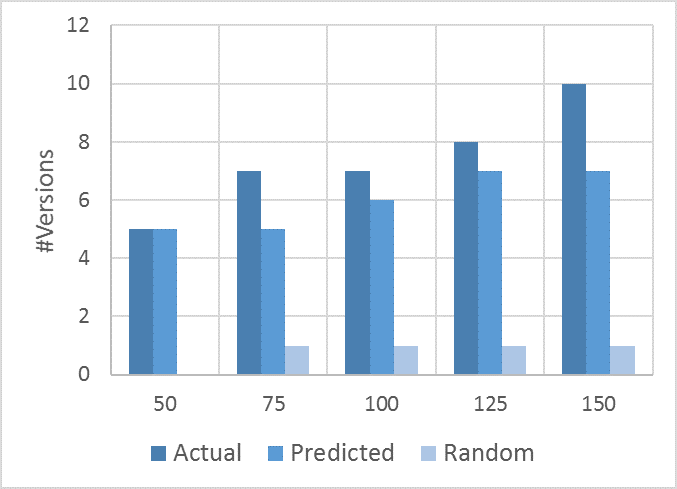}
 \caption{Math project. \# of versions that converged for different test budgets.} 
 \label{fig:bars-math}
\end{figure}

\subsubsection{Results}  Figures~\ref{fig:bars-lang} and~\ref{fig:bars-math} plots the number of bugs in which each algorithm has converged, for $B=50, 75, 100, 125$, and $150$, for the Lang and Math projects, respectively.  The results clearly show that Random fails to converge much more often, compared to Predicted and Oracle. For example, in the Lang project with a test budget of 150, Random converges in fewer than 10 versions while both Oracle and Predicted converged in more than 50 versions.

%both projects and for every stopping condition the gray bar is significantly lower than both blue and orange. 
When comparing Prediction and Oracle, we observe very similar results in most cases, with a small advantage for Oracle. %For example, can see that for all tests budgets except to 75 the oracle manage to converged in more cases than the predicted algorithm.
The advantage of Oracle is expected because it uses the actual test traces to choose which test to execute. However, this advantage is very small, suggesting that our trace prediction algorithm works well for guiding tests in a software  troubleshooting process. 

Note that the advantage of Oracle over Predicted is larger for Math than for Lang. This corresponds to the trace prediction accuracy results reported in the previous section, as the AUC for Lang is significantly higher than the AUC for Math. This shows the relation between the strength of the trace prediction and the effectiveness of the troubleshooting.

% Another perspective to observe the results is by checking how many time every algorithm stops because of reaching the first stopping condition (score 0.7) and didn't stop because of tests limitation.As we can see in the following table:

% Table~\ref{tab:budget} presents the results of our troubleshooting experiments, when we changed the number of tests limitation for every experiment.
% We provide below a brief description of each of the columns. 
% \begin{itemize}
%  \item \textbf{Budget}-Number of tests each algorithm can run before we stop it.
%  \item \textbf{Bugs that converged}-in how many versions the algorithm stop because it reaches the score of 0.7.For example, in Lang project the algorithm that uses the real traces manage to reach the 0.7 score in 53 bugs.
%  \item \textbf{Avg. pre}- The average final precision calculated over all the bugs. Ie, we recorded the final precision for every bug and algorithm and calculate it's average. 
%  \item \textbf{Avg. recall}- The average final recall calculated over all the bugs. Ie, we recorded the final recall for every bug and algorithm and calculate it's average. 
 
%  \end{itemize}

%\Roni{TODO-EYAL: Change these figures to the 150 results}Eyal - Done, just change the figures sizes.

\begin{figure}
 \centering
  \includegraphics[width=0.85\columnwidth]{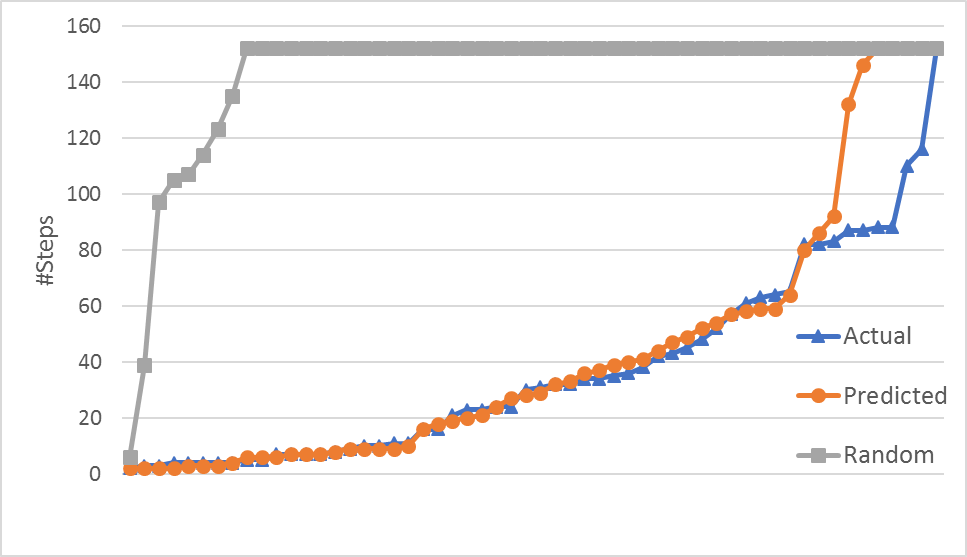}
 \caption{Lang project. \# of steps until reached one of the stopping conditions.} 
 \label{fig:cactus-lang}
\end{figure}

\begin{figure}
\centering
  \includegraphics[width=0.85\columnwidth]{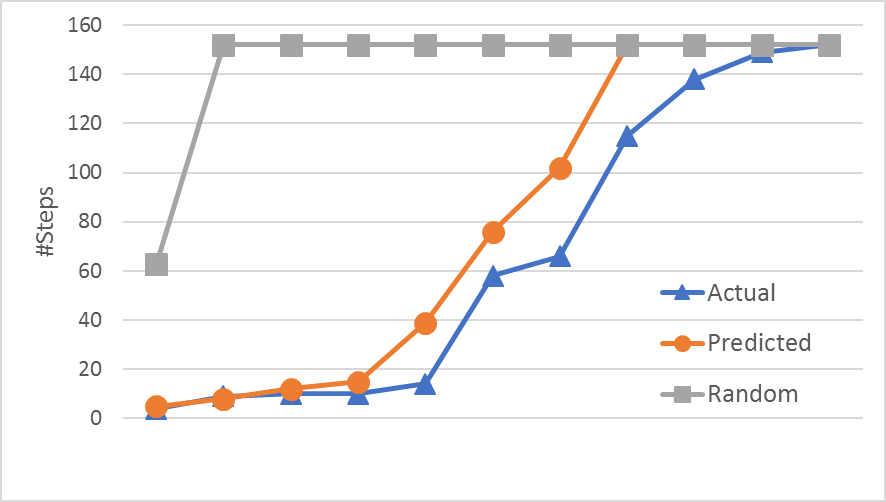}
 \caption{Math project. \# of steps until reached one of the stopping conditions.} 
 \label{fig:cactus-math}
\end{figure}

To gain a deeper insight into our results, Figures~\ref{fig:cactus-lang} and~\ref{fig:cactus-math} presents the number of steps until each algorithm converged for the cases where it did not reach a timeout. 
The $y$ axis represents the number of steps performed, for different versions and \tp algorithms. Every line represents a different \tp algorithm and every data point represents a single experiment, i.e., the number of steps until convergence for a specific bug and \tp  algorithm. The data points for every test planning algorithms are sorted on the $x$-axis according to their number of steps. %For example, in the Math project when using Random, only we manage to reach diagnose with score higher than 0.7 just one time, because in the rest of the bugs it required to run the maximum of 150 tests. TODO: FIND A BETTER EXAMPLE. 
%\Roni{TODO-EYAL: Update the example in the above sentence with the new data}Eyal - Done.
Of course, being lower on the $y$-axis is better, as it shows fewer steps are need to coverage. In this figure, we did not show bugs for which all algorithms reached the budget without converging. 

% \begin{figure*}
%   \includegraphics[width=0.48\linewidth]{Images/lang_150_steps.png}
%   \includegraphics[width=0.45\linewidth]{Images/Math_150_steps.png}
%  \caption{Number of steps until reached one of the stopping conditions for Lang (left) and Math (right) projects.} 
%  \label{fig:cactus}

% \end{figure*}

Similar trends are observed in these results. Random performs significant worst than Predicted and Oracle, in both projects. Again, Oracle performs the same or better than Prediction in almost all cases. However, the results clearly show that the difference between them is not large. 

%We can also see that there is a direct connection between the classifier results to the troubleshooting results. When learning, Lang project had better results that Math. and it reflected in the troubleshooting results as well.For example, when comparing the two graphs we can see that at most cases in Lang project the Prediction and Oracle are the same (except of the 4 last points the lines are almost identical). in contrast to the Math project where we can see that in all version except one the Oracle manage to converged after fewer step than the Prediction.\Roni{TODO EYal, try to re-phrase this}Eyal - I tried, and move the correlation between learning and troubleshooting here.

%\Roni{TODO: Summary}

\section{Related Work}

Tracing the dynamic execution of a program has many applications, and consequently, a large body of research is devoted to study different aspects of efficient and effective tracing. To the best of our knowledge, no prior work attempted to learn how to predict a trace of a test by analyzing traces of other tests. The most similar research we have found is Daniel and Boshernitsan~\cite{Daniel2008PredictingEO} work on predicting the effectiveness of test generation algorithms. They propose to use supervised learning to train a classifier that predicts the coverage of a given function obtained by running a given test. The label they aimed to predict is how much the function is covered, while we aim to predict the actual trace. Thus, our trace prediction can be used to estimate coverage, but not vice versa. An interesting future work would be to use our trace prediction algorithm to estimate coverage.

One branch of research studied different ways to visualize collected traces in a meaningful way~\cite{de2002visualizing,cornelissen2007understanding}. 
A different branch aims to reduce the size of the generated traces, e.g., by compressing them on-the-fly~\cite{reiss2001encoding,taheri2017parlot} or by identifying the parts of the code that are not relevant for the task at hand and thus can be removed from the trace (or not traced to begin with)~\cite{tallam2007enabling} 

Other prior work aimed to minimize the cost of tracing by choosing intelligently which components to monitor, consider this as a sensor-placement problem~\cite{zhao2017log20}. 
Liblit et al.~\cite{liblit2003bug,liblit2005scalable} proposed a statistical method to sample traces, in the context of bug isolation. Their approach is somewhat similar to our work, but in their task they are given a part of the trace. Thus, they must execute the test while our trace prediction algorithm does not. % and aim to statistically complete it while we have  trace of the program to as they tried to predict the parts of the trace that were not sampled. However, our trace prediction problem is more difficult, in that we have no knowledge about the trace do not have any sampled component. These work differ from our work, as we do not sample some of the components in the trace but fully predict it based on 

% [[Roni: this is all about test generation. How is it related?]] Our research is related to several lines or research. 
% Our research is come to improve the software testing process, process that is well studied. The focus so far in this field was to write and execute tests with high code coverage as possible, especially generating those tests automatically instead of writing them manually as it is well explained in those researches~\cite{pacheco2007randoop,tillmann2008pex,zhang2010test}. 
% Code coverage in general is a broad subject that is also related to automatic test generation. Since the assumption that a higher code coverage is better, one of the main goals in automatically tests generations is to achieve as much coverage as possible. In a Google study that focus on how to best use an automated unit test generation tool to maximize code coverage~\cite{campos2015continuous}. The researchers developed a tool that automatically generates unit tests for Java software and used an evolutionary algorithm to generate them.
% In contrast to our case where the goal is to select 
% what test to run each time, so a test will be considered good if it leads to bug identification.

\subsection{Threats to Validity and Limitations}
Our approach for trace prediction requires a large set of automated tests to be available
and executable in past versions of the code. Thus, our work is not suitable for projects in an early stage. Future work may study how to learn  trace prediction from one project to the other. 
The main limitation of this research is the breadth of our experimental evaluation. Future work will perform a large-scale study over more projects and more programming languages.

\section{Conclusion and Future Work}

We proposed a trace prediction algorithm that learns from a small fraction of existing traces how to predict whether a given software component is in a trace of a given test. Our approach is based on modeling the trace prediction problem as a binary classification problem and applying supervised learning to solve it. To this end, we propose features based on call graph analysis and syntactic similarity, and show experimentally that they work well on two open-source projects. 
Then, we show how predicted traces can be used in a \tp algorithm, as part of \LDP, a recently proposed software troubleshooting paradigm. We evaluate \LDP without \tp algorithm, showing that it converges much faster than random test selection, and almost the same as an Oracle \tp algorithm, that knows a-priori the tests' traces. 

While our results are encouraging, there is much to do in future work. First, 
the quality of our trace prediction is far from perfect. This is because the available data is highly imbalanced, where the vast majority of component-test pairs are negative (i.e, the component is not in the trace of the test). Future work can attempt to address this using known techniques for imbalanced dataset. In addition, future work will extend our experimental evaluation to more projects, and include a user case study. Another interesting direction is to combine our trace prediction algorithm with symbolic execution methods, as well as strongly \tp algorithms. %standard imbalance technqieu trace prediction  The prediction problem is especially challenging as the dataset is highly imbalanced. Nevertheless, with fairly  highly imbalanced. NThe results show that the trace prediction algorithm is 

\section{Acknowledgements}
This research was partially funded by the Israeli Science Foundation (ISF) grant \#210/17 to Roni Stern. 
\bibliographystyle{IEEEtran}
\bibliography{library}
\end{document}